\documentclass[a4paper,11pt]{article}
\pdfoutput=1

\usepackage{jheppub}
\usepackage[T1]{fontenc}
\usepackage{amsmath,amsfonts,amssymb,textcomp}
\usepackage{slashed}
\usepackage{graphicx}
\usepackage[dvipsnames,table]{xcolor}
\usepackage{booktabs}

\def\beq{\begin{equation}}
\def\eeq{\end{equation}}
\def\beqa{\begin{eqnarray}}
\def\eeqa{\end{eqnarray}}

\def\ltap{\ \raise.3ex\hbox{$<$\kern-.75em\lower1ex\hbox{$\sim$}}\ }
\def\gtap{\ \raise.3ex\hbox{$>$\kern-.75em\lower1ex\hbox{$\sim$}}\ }
\def\ba{\begin{array}}
\def\ea{\end{array}}
\def\bea{\begin{eqnarray}}
\def\eea{\end{eqnarray}}
\def\bean{\begin{eqnarray*}}
\def\eean{\end{eqnarray*}}

\def\cL{{\cal L}}
\def\cA{{\cal A}}

\def\tr{{\rm Tr}}

\usepackage[normalem]{ulem}

\newcommand{\cQ}{{\cal Q}}
\newcommand{\gZp}{g_{Z'}}
\newcommand{\mZp}{m_{Z'}}

\subheader{\footnotesize
\vspace*{-3.7em}
\begin{flushright}
CERN-TH-2017-254\\
PITT-PACC-1716
\end{flushright}
}

\title{Anomalous $Z'$ and Diboson Resonances at the LHC}

\author[a]{Ahmed Ismail}
\author[b,c]{and Andrey Katz}

\affiliation[a]{Pittsburgh Particle Physics, Astrophysics, and Cosmology Center,\\Department of Physics and Astronomy, University of Pittsburgh, Pittsburgh, USA}
\affiliation[b]{Theory Division, CERN, CH-1211 Geneva 23, Switzerland}
\affiliation[c]{D\'{e}partement de Physique Th\'{e}orique and Center for Astroparticle Physics (CAP),\\Universit\'{e} de Gen\`{e}ve, 24 quai Ansermet, CH-1211 Gen\`{e}ve 4, Switzerland}

\emailAdd{aismail@pitt.edu}
\emailAdd{andrey.katz@cern.ch}

\abstract{ We propose novel collider searches which can significantly improve the LHC reach to new gauge bosons $Z'$ with 
	mixed anomalies with the electroweak (EW) gauge group. Such a $Z'$ necessarily acquires a Chern-Simons 
	coupling to the EW gauge bosons and these couplings can drive both exotic $Z$ decays into $Z'\gamma$ if the 
	new gauge boson is sufficiently light, as well as $Z'$ decays into EW gauge bosons.
	While the exotic decay rate of the heavy $Z$ into $Z'\gamma$ is too small to 
		be observed at the LHC,
	for a light $Z'$, we show the potential of a lepton jet search in association with a photon to probe the rare decay $Z \to Z' \gamma$. 
}

\begin{document} 
\maketitle
\flushbottom

\section{Introduction}
\label{sec:intro}

The motivations for new $Z'$ gauge bosons, both heavier and lighter than the EW 
scale, are ubiquitous~\cite{Fayet:1990wx,Langacker:2008yv}.  These theoretical expectations stimulated 
multiple experimental searches at the LHC as well as earlier colliders including
LEP and the Tevatron, in addition to B-factories, fixed target experiments and 
other facilities. 

 New $Z'$s lighter than the mass of the SM $Z$-boson (that we will dub ``light $Z'$s'') 
 have been proposed in the context of theories 
of light dark matter (DM)~\cite{Essig:2009nc,Lin:2011gj}, various attempts to explain perceived anomalies in data, 
{\it e.g.} the 511~keV line~\cite{Boehm:2003ha}, the muon anomalous magnetic dipole moment~\cite{Gninenko:2001hx}, the proton charge radius~\cite{Barger:2010aj,TuckerSmith:2010ra}, flavor measurements~\cite{Altmannshofer:2013foa,Gauld:2013qba}
and many others.  While the strengths of particular motivations for this type of scenario might 
depend on one's taste, a light and weakly coupled 
$Z'$ is a necessary ingredient of many realistic BSM scenarios, and deserves
careful dedicated studies. 

Currently most experimental bounds on the light $Z'$ scenario arose 
from low energy experiments, LEP and astrophysical observations
(see~\cite{Essig:2013lka} for a comprehensive review). In particular, to probe this particle 
one would rely on exotic flavor-changing neutral current (FCNC) processes, 
measurements of the electron and muon magnetic 
dipole moments, beam dump experiments and energy emission 
observation by supernovae. In the more massive region of the parameter space, {\it i.e.} between the 
MeV scale and the $Z$ mass, the former two classes of constraints dominate the exclusions, 
while beam dump experiments are important only in the case of $Z'$s that are 
extremely feebly coupled to the SM, such that their displaced decays can be observed. 
Therefore, the constraints that one usually quotes as the dominant constraints on the light $Z'$s
are mostly due to the \emph{leptonic couplings of the $Z'$}, with a noticeable exception of the 
rare meson processes.

Because the couplings of the $Z'$ are highly model dependent, constraints on new gauge bosons
are often phrased in terms of the benchmark dark photon model, where the only parameters are
the $Z'$ mass and its kinetic mixing with the photon. One can 
argue on general grounds that for any $Z'$ which is not completely sequestered from the 
SM, this kind of kinetic mixing will be present, and even if it is absent at tree level, will necessarily
be formed radiatively~\cite{Holdom:1985ag}. 
While this approach is relatively generic, it implicitly makes an extremely 
important assumption, that SM matter is not directly charged under the force mediated 
by the light $Z'$. Clearly, many models that we have listed above do not satisfy this assumption. 

In this work we focus on a broad class of models, in which at least some of the SM fermions 
are charged under the dark $U(1)'$. Of course, if we assume no extra fermions charged under the 
SM and the dark $U(1)'$, the choice of allowed Abelian symmetries is strictly limited
to linear combinations of the 
$B-L$, $Y$-sequential, and inter-generational symmetries. However, it is possible, and, in 
fact, desirable to extend the scope of $Z'$ searches to theories that appear to be \emph{anomalous}
with the field content of the SM. As has been emphasized 
long ago~\cite{D'Hoker:1984ph,D'Hoker:1984pc,Preskill:1990fr}, these naively anomalous theories
can be formulated as Effective Field Theories (EFTs) well below the scale of the masses of the 
exotic fermions, the ``spectators''. Systematic integration out of the spectators 
yields Chern-Simons terms. These terms naively appear to be renormalizable, yet correspond to
derivative couplings of longitudinal gauge bosons. They are 
not gauge invariant, and below the EFT cutoff produce amplitudes that grow with energy. 
Recently these kind of effective theories have been studied in detail both in the context of 
anomalous DM mediators~\cite{Ismail:2017ulg}, and in the context of 
light $Z'$s~\cite{Dror:2017ehi,Dror:2017nsg}.   

In this paper we point out that due to the Chern-Simons terms,
light anomalous $Z'$s can induce new distinctive LHC signatures, that can potentially 
significantly improve the reach, compared to existing constraints. Most noticeably, 
the Chern-Simons terms induce the exotic $Z$ decays, $Z \to Z' \gamma$.  This decay mode, 
followed by a subsequent leptonic $Z'$ decay can be a 
spectacular signature even at a 
hadron collider. 

We will further concentrate on the light $Z'$ in the mass range between 1~GeV and the 
$Z$ boson mass. 
We demonstrate 
that the LHC, and especially the high luminosity LHC (HL-LHC) can 
potentially probe $U(1)'$ couplings at or even below the $10^{-2}$ level  
via searches for these exotic $Z$ decays. 
As we will later see, this $\gZp \sim 10^{-2}$--$10^{-3}$ reach is comparable to the 
LHC reach 
to similar scenarios due to the direct production of the $Z'$~\cite{Hoenig:2014dsa}
as well as to the LEP and Tevatron 
sensitivities to similar scenarios~\cite{Appelquist:2002mw, Carena:2004xs}. 
We emphasize however
that the search that we propose is complementary to the existing searches, because 
exotic $Z$ decays searches can also  
probe $U(1)'$ symmetries that do not couple to the first generation SM fermions at all 
and which are therefore inaccessible to existent direct production searches.  

Our paper is organized as follows. In the next section we will briefly review the theoretical background, calculate the exotic BR of the $Z$ boson and present several 
simplified benchmark models, that can be probed by the techniques that we propose.
In Sec.~\ref{sec:constraints} we review the existing constraints on light $Z'$s that come 
from direct searches in earlier colliders and $B$ factories (including BaBar, KLOE and LEP), exotic 
decays of $B$ mesons,  beam dump experiments and astrophysical observations. We describe 
the search that we propose and estimate its potential reach in Sec.~\ref{sec:exoticdecays}.
Finally, in the last section we conclude.

\section{Benchmark Models}
\label{sec:models}

\subsection{Theory background}
Consider a $U(1)'$ symmetry that has a mixed anomaly with the EW group $SU(2)\times U(1)$. Clearly 
this theory cannot be extended up to arbitrary high scales~\cite{Preskill:1990fr}. 
However, it can be formulated as a low 
energy effective theory with a cutoff 
\beq\label{eq:cutoff}
\Lambda \lesssim \frac{64 \pi^3 \mZp}{3 \gZp g_{SM}^2}~,
\eeq  
where $g_{SM}$ is usually the largest coupling of the SM gauge group with which 
the $U(1)'$ has an anomaly, unless the effective theory has a very big anomaly coefficient. 
The full theory should be augmented with the ``spectators'', the new fermions, charged both 
under the SM and the $U(1)'$ that eventually cancel the anomaly caused by the SM fermions. 
However, the masses of the ``spectators''
can be as heavy as $\Lambda$ without compromising the validity of the theory. 
Note also that we assume that these fermions are chiral under the $U(1)'$ 
and vector-like under the SM, thus getting their masses from the $U(1)'$ 
breaking. While the opposite pattern is possible conceptually, 
it is very challenging to reconcile with experimental bounds on
new fermions that are chiral under the SM.

We will further assume that the ``spectators'' are beyond the experimental reach of the 
colliders, whether the $Z'$ is heavy or light, and integrate them out. 
After integrating them out one finds that the anomalous EFT possesses new Chern-Simons terms, that couple the 
$Z'$ to the gauge fields of the groups with which it has a mixed anomaly:
\beq\label{eq:WZterms}
\cL  \sim  C_B \epsilon^{\mu \nu \rho \sigma } Z'_{\mu} B_\nu \partial_\rho 
B_\sigma + 
C_W
\epsilon^{\mu \nu \rho\sigma} Z'_{\mu} \left(W^a_{\nu} \partial_\rho W^a_{\sigma} +\frac{1}{3} \epsilon^{abc} 
W^a_\nu W^b_\rho W^c_\sigma \right)
\eeq

Strictly speaking the sizes of the counterterms corresponding to $C_B$ and $C_W$ can be
completely arbitrary and depend on the 
momentum shift between the triangle loop diagrams when a full matrix element is calculated
(see~\cite{Preskill:1983} for a detailed explanation
and~\cite{Racioppi:2009yxa,Anastasopoulos:2008jt,Ismail:2017ulg} for more recent 
overviews). However, for the purposes of the calculations there are two different gauges that 
turn out to be particularly useful. In one of them, one can simply set the counterterm coefficients $C_B$
and $C_W$ to zero and absorb all of the anomaly effects into the momentum shift between the diagrams.
These momentum shifts, in turn, are chosen to satisfy the Ward identities for the anomaly-free gauge groups, $SU(2) \times U(1)_Y$. This is the approach that we have recently adopted
in~\cite{Ismail:2017ulg}. 

In this paper we find another approach to be particularly useful.
 We choose a gauge such that there is 
no contribution from the 
momentum shift between the anomaly triangle diagrams, which 
then exactly cancel one another.   
In this case, to preserve $SU(2) \times U(1)_Y$ gauge invariance the counterterms of~\eqref{eq:WZterms} are given by
\beq\label{eq:ccoeff}
C_B = \frac{\cA_{Z'BB}}{12 \pi^2} \gZp g^2, \ \ \ \ C_W = \frac{\cA_{Z'WW}}{12 \pi^2} \gZp {g'}^2~,
\eeq
with the anomaly coefficients 
\beq\label{eq:acoeff}
\cA_{Z'BB} = \tr (Q' Y^2), \ \ \ \ \cA_{Z'WW} = \tr (Q' T^a T^a)~. 
\eeq
With either of the gauge choices above, the 
physical observables are of course unchanged. However, the latter prescription fully captures the effect of the spectator masses in the Chern-Simons terms, and is simple to implement in the calculations here. 

Let us start first from a light $Z'$ and consider a situation 
in which the exotic decays of the SM $Z$ are allowed. 
Starting from the couplings~\eqref{eq:ccoeff}
and~\eqref{eq:acoeff} 
it is straightforward to project the gauge fields 
$W^3_\mu$ and $B_\mu$ onto the mass eigenstates and calculate the $Z$ exotic 
decay width into $Z'\gamma$. One finds, assuming $m_{Z'} \ll m_Z$:\footnote{Note the factor of 
9 discrepancy in the denominator with Ref~\cite{Dror:2017nsg} due to the slightly different definitions of the 
$\cA_{Z'BB},\ \cA_{Z'WW}$ coefficients.}
\beq\label{eq:Zexoticwidth}
\Gamma(Z \to Z' \gamma) =  \frac{|{g'}^2 \cA_{Z'BB}  - g^2 \cA_{Z'WW} 
	|^2}{3456 \pi^5} \gZp^2\sin^2\theta_W\cos^2 \theta_W \frac{m_Z^3}{\mZp^2}~.
\eeq 

As expected this expression is enhanced by the ratio $(m_Z/\mZp)^2$, 
manifestly signaling that 
the theory is not renormalizable and that the coupling between the gauge bosons grows with the energy. 

\begin{figure}
	\centering
	\includegraphics[width=.6\textwidth]{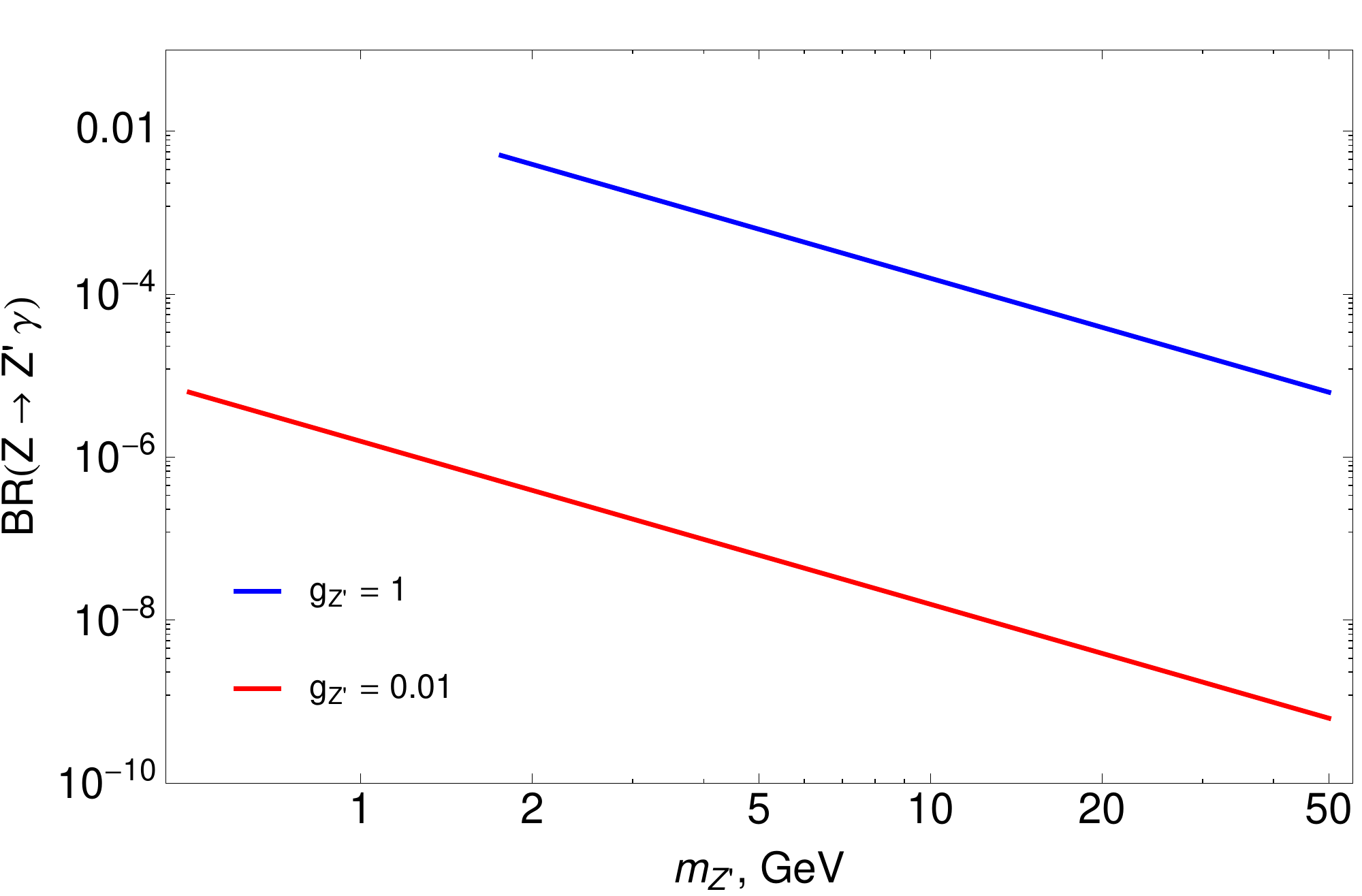}
	\caption{The exotic BR($Z \to Z' \gamma$) as a function of the $Z'$ mass.
	 We assume $\gZp = 1$ for the 
	blue line and $\gZp = 0.01$ for the red line. The anomaly coefficients are chosen to be $\cA_{Z'BB}
	= -\cA_{Z'WW}$ = 1. We assume the cutoff of the theory~\eqref{eq:cutoff} to be $\Lambda = 5$~TeV
    and draw the BR line only in the region where the EFT is well defined.}
\label{fig:exoticZBR}
\end{figure}

We show the exotic $Z$ BR in Fig.~\ref{fig:exoticZBR}. There for simplicity we assume 
$\cA_{Z'BB} = -\cA_{Z'WW} =1$, where the first equality is expected to be true in any vector-like $U(1)'$, such that there is no 
mixed anomaly with $U(1)_{EM}$ or the SM 
color group $SU(3)_c$. In order to define the range of validity of the theory we 
should also address the issue of the cutoff as defined in~\eqref{eq:cutoff}. 
Hereafter we assume the cutoff of 5~TeV, making sure that 
the spectators can be safely pushed to scales that are inaccessible to the LHC. 
Because of 
the enhancement at low $Z'$ masses, 
the main gains of our proposed LHC searches will be in the light $Z'$ case.   

For completeness we also consider here the case in the which the $Z'$ is heavier than 
the mass of the $Z$. In this case it is straightforward to calculate the decay width 
of the $Z'$ into $Z\gamma$.
This expression, assuming $m_{Z'} \gg m_Z$, is qualitatively different
from~\eqref{eq:Zexoticwidth}:
\beq
\Gamma(Z' \to Z\gamma) =  \frac{|{g'}^2 \cA_{Z'BB}  - g^2 \cA_{Z'WW} 
	|^2}{13824 \pi^5} \gZp^2\sin^2\theta_W\cos^2 \theta_W \frac{m_Z^2}{\mZp}~.
\eeq   
Unlike~\eqref{eq:Zexoticwidth} this expression is in fact suppressed by the lighter $Z$ mass. This behaviour occurs because the effects of the integrated out spectators are suppressed by the cutoff of the theory,
which is proportional to the $Z'$ mass~\eqref{eq:cutoff} due to the spectators being chiral under $U(1)'$. This renders the 
exotic width of the heavy $Z'$ into $Z\gamma$ practically unobservable and therefore 
we do not pursue this direction any further.

\subsection{Light $Z'$ Models}  
Here we survey models that could have relevant signatures for a 
$Z'$ that is lighter than the mass of the SM $Z$.

While the LHC produces an immense quantity of $Z$ bosons---at $\sqrt{s} = 13$~TeV 
we have 
$\sigma (Z + X ) \approx 60$~nb~\cite{Aad:2016naf}---we 
are interested in very small BRs with non-trivial background. Therefore, we will further concentrate 
on the theories, where the $Z'$ has appreciable leptonic BRs and consequently the associated $U(1)'$ will be related to lepton number. In addition, as explained in 
the introduction, we will be interested in theories that have mixed anomalies with the SM 
$SU(2)\times U(1)_Y$. Here we describe a handful of symmetries that we have in mind for these searches.    

\paragraph{Lepton number.}
A classical example of a theory along these lines would be a lepton number gauge symmetry. 
Under this symmetry we give positive charge to the left-handed leptons and negative charge to the right-handed ones: $Q(L) = -Q(e^c) = 1$. Of course the neutrino mass 
operator violates this symmetry, suggesting a relatively low energy mechanism for the neutrino 
masses (below the EFT cutoff $\Lambda$). However, scenarios along these lines have been proposed and we believe 
that neutrino
masses do not pose conceptual problems to the gauged lepton number.

One of the main advantages of this symmetry is that all of the SM Yukawa couplings (though not
the Weinberg operator) respect this symmetry even if the SM Higgs is uncharged under $U(1)'$. Consequently there is no mixing between the $Z$ and $Z'$, and 
EW precision measurements are not affected. As expected in all theories that keep the Yukawa 
couplings gauge invariant, this symmetry has no mixed anomaly with $SU(3)_c \times U(1)_{EM}$. 
The anomaly coefficients are given by $\cA_{Z'BB} = - \cA_{Z'WW} = - N_g/2$, where $N_g$ stands for the 
number of generations. 

The dominant constraints on such a leptophilic $Z'$ come from direct production and exotic $Z$ decays at LEP; the former is typically much stronger. 
For lighter $Z'$s with mass $\lesssim 10$~GeV, one is also
subject to stringent constraints from direct 
production at KLOE~\cite{Anastasi:2015qla} and BaBar~\cite{Fayet:2007ua}. 
The second class of constraints on this 
scenario is exotic heavy meson decays, 
similar to those for light mesons that were first emphasized in~\cite{Tulin:2014tya}. 
While naively it would seem that we should have no 
coupling between the $Z'$ and hadrons, because we postulate a lepton number symmetry, 
this kind of coupling does emerge due to the kinetic mixing between the $Z'$ and hypercharge. 
A priori we do not know how big this coupling is, however on naturalness grounds we expect 
it to be at least of order 
\beq\label{eq:KineticMix}
\kappa \sim \frac{g' \gZp}{16\pi^2}  \log \left( \frac{ M}{m_l}\right)~,
\eeq
where $M$ is some high scale where this mixing is formed, and $m_l$
is the mass of the lightest fermions, that are charged both under the the $U(1)'$ and 
the hypercharge.  

Finally let us note that in our subsequent analysis we will also address the gauging of a lepton number 
symmetry that acts only on one or two generations (e.g. muon number or muon-plus-tau
number). While there is no conceptual problem in defining these symmetries and they do not 
lead to FCNCs in 
isolation,\footnote{Through its anomalies with $SU(2)$, 
	such a $Z'$ can still contribute to FCNCs mediated by $W$ loops and proportional to the CKM matrix elements. 
	We will review these processes in the next section.} these symmetries will be especially interesting in the context of our study because 
one cannot produce the $Z'$ directly in $e^+e^-$ collisions  (except for the kinetic mixing-suppressed production), and therefore exotic $Z$ decays are one of the only accessible ways to 
probe these models experimentally.    

\paragraph{$B+L$ symmetry.} We introduce this symmetry because both the leptons and the 
hadrons are charged under it and therefore it is subject both to LEP and LHC 
direct production constraints.  We assign the $U(1)'$ charges as follows:
\beq\label{eq:b+l}
Q(L_i) = -Q(e^c_i) = 1, \ \ \ \ \ Q(\cQ_i) = -Q(u^c_i) = -Q(d^c_i) = 1
\eeq

Exactly as in the previous scenario, we will further 
consider a $B+L$ symmetry that acts on all the generations, along with a $B+L$ model
that acts only on the heavy flavors, thus avoiding the most stringent constraints from direct searches. 
Like the previous symmetry it is also vector-like and therefore
does not have any mixed anomalies with the $SU(3)_c \times U(1)_{EM}$ subgroup of the SM. The anomaly coefficients 
for this group are $\cA_{Z'BB} = $\, $\cA_{Z'WW} = - 2 N_g $. Finally we notice that we expect the 
kinetic mixing between the hypercharge and the $U(1)'$ symmetry to be the same as~\eqref{eq:KineticMix}
except that instead of $m_l$ in the logarithm we can have either the lepton or the quark mass, whichever
is lighter. However, unlike in the previous scenario, this mixing does not have any important consequences
on the detectability of the $Z'$.

\paragraph{Right-handed lepton number.} This is the only model that we consider that is not 
vector-like under $U(1)_{EM}$ and therefore contains the anomalous coupling $\gamma \gamma Z'$
does not vanish (although the decay rate $Z' \to \gamma \gamma$ still vanishes due to the Landau-Yang
theorem). We assign (some or all of) the right-handed leptons $e^c_i$ charge $+1$, while leaving the rest
of the SM fermions uncharged. The only non-vanishing anomaly coefficient is $\cA_{Z'BB} = -N_g$, 
and the minimal mixing with hypercharge in the absence of fine-tuning is expected to be 
identical to~\eqref{eq:KineticMix}.   It is also worth emphasizing, that because this symmetry does not have a 
mixed anomaly with $SU(2)_L$, it evades some flavor physics bounds that we discuss in the next section. 


\section{Existing Constraints on Light Z's and the LHC Searches}
\label{sec:constraints}

While the experimental bounds on a heavy $Z'$ are in most 
cases straightforward and set by LHC resonance searches, 
which by now largely supersede older LEP and Tevatron bounds, 
the situation with a $Z'$s lighter than the 
SM $Z$ is more subtle.     
A light $Z'$ has been previously searched for at various 
collider experiments, including LEP, BaBar and the LHC.
As we will see, most of the existing bounds that are relevant for the scenarios that we have 
outlined in Sec.~\ref{sec:models} are still coming from these searches. 
In this section we will
review these bounds, as well as various other constraints that arise from 
other searches, most notably flavor measurements.

In the low mass range, the strongest bounds are coming from the 
BaBar 
experiment, and, for masses
below 1~GeV, from KLOE. The latter mostly 
searched for a  
$Z'$ with flavor-blind couplings to the SM, 
motivated by a hidden $U(1)'$ under which the 
SM particles are not directly charged, so that the couplings come from kinetic mixing between the 
SM photon and the $Z'$. 
The relevant searches of KLOE for a $Z'$ decaying into 
electrons~\cite{Anastasi:2015qla}  (that excludes the $Z'$ masses below 
500~MeV, with the exclusions often not exceeding those of NA48/2~\cite{Batley:2015lha})
and charged pions~\cite{Anastasi:2016ktq} constrain the kinetic mixing parameter 
$\kappa \lesssim 10^{-3}$ in the mass range between 500~MeV and 1~GeV. As we have noted, 
we will be interested in the $Z'$ mass range above 1~GeV because 
of the practical difficulty in resolving resonances lighter than 1~GeV at the LHC,   
and, therefore, these
constraints are of limited interest for us. 

The mass range between 1 and 10~GeV currently 
enjoys the best coverage from 
the BaBar experiment. BaBar
has both searches for a $Z'$ that couples directly to 
electrons~\cite{Lees:2014xha}, and for a 
$Z'$ that only couples to the second and third 
generations of leptons~\cite{TheBABAR:2016rlg}.  In the 
former case BaBar looks for $e^+ e^- \to \gamma Z'$ 
events with a subsequent decay of the 
$Z'$ into a pair of electrons or muons. The bounds are quoted by BaBar in terms of the kinetic 
mixing $\kappa \lesssim 10^{-3}$--$10^{-4}$ assuming that the coupling between the SM fermions and 
the $Z'$ is merely due to the kinetic mixing. If we interpret these results in terms of the 
lepton number gauge boson, such that the electron and muon are directly charged under the 
hidden $U(1)'$, we constrain the gauge coupling 
$\gZp \lesssim 10^{-4}$.  
We show these constraints explicitly on Fig.~\ref{fig:searchreach} together with 
the other relevant bounds. 
These are the dominant constraints on the 
leptophilic $Z'$ in the range between 1~and 6~GeV. This 
search for a universal leptophilic $Z'$ is extremely 
robust and the bounds are unlikely to be improved at the LHC. 

We note in passing that for an extremely light leptophilic $Z'$, there are additional 
constraints from neutrino-electron scattering that could be stronger than those that we have mentioned already below $m_{Z'} \sim 1$~GeV~\cite{Bilmis:2015lja,Jeong:2015bbi}. Given our mass region of interest, we do not show these bounds in what follows.

The constraints are naturally more modest if we assume that 
only the second and third generation 
SM fermions are charged under the hidden $U(1)'$. 
This scenario has been analyzed by BaBar  
in Ref.~\cite{TheBABAR:2016rlg} by considering the process $e^+ e^- \to \mu^+ \mu^- Z'$, 
where 
the $Z'$ 
is radiated off of a muon.
Even though these constraints are slightly weaker than 
those that one gets in the case of the direct coupling to the electron, 
they are still extremely 
strong, excluding couplings of order $\gZp \lesssim 10^{-2}$ and even 
slightly smaller. We also show these bounds 
on Fig.~\ref{fig:searchreach} together with the projected reach of the proposed LHC searches. 
 As we will further see, this bound can be significantly improved by the search 
 for exotic $Z$ decays that that we propose 
 in Sec.~\ref{sec:exoticdecays}.
 
 As expected, the direct searches from BaBar cannot be efficient above masses 
 of approximately 6--7~GeV, and we are forced to switch to other experiments. 
 Searches at LEP, which would produce the $Z'$ in conjunction with the photon, give a 
 generic bound around $\gZp \lesssim 10^{-2}$~\cite{Appelquist:2002mw}, provided that the mass 
 of the $Z'$ is not too close to the center-of-mass energy of any of the LEP runs, namely 
 $\sqrt{s} =130, 136, 161, 172, 183 $~GeV, which is always true in the case of the light 
 $Z'$. 
 However, as we will immediately see, the 
 existing LHC searches already outperform these results.
 
 Unfortunately there is no dedicated search for a
  light $Z'$ at ATLAS or CMS, except in the leptophobic case where a limit of
  the order of $\gZp \lesssim 10^{-1}$ has recently been set~\cite{Sirunyan:2017nvi}.
  However, a recast in the 
 context of direct $Z'$ production in the Drell-Yan process has been 
 performed in~\cite{Hoenig:2014dsa}. The main conclusion of this reference is that 
 the bounds that CMS puts on a light $Z'$ from direct searches are 
 meaningful and already improve on the existing LEP bounds.\footnote{The ATLAS 
 data	 
 were not explicitly recasted, though they can arguably yield similar 
constraints. } 
  Ref.~\cite{Hoenig:2014dsa}
 interprets these bounds again in terms of the
 kinetic mixing between the visible and the hidden photon. The bound that these searches 
 put is around $\kappa \lesssim 10^{-2}$ in the range of masses between 10 and 30~GeV,
  rising to a few$\times 10^{-2}$ in the range between 30 and 80~GeV.
 This would 
 impose a rather strong bound on a \emph{lepton-baryon universal model} 
 	such as $U(1)_{B+L}$, of order $\gZp \lesssim 10^{-3}$; however, the bounds on 
 leptophilic models are much more modest and, assuming the 
 kinetic mixing as in~\eqref{eq:KineticMix}, does not even exceed 
 $\gZp \sim 0.1$, weaker than LEP. 
 
Another important observation in the context of the LHC searches has been 
made in~\cite{Altmannshofer:2014pba}. Once we have a light $Z'$ that couples to the 
SM leptons (not necessarily the electrons), it can be radiated off of the final state in the 
decay $Z\to \ell^+ \ell^-$, leading to an exotic $Z$ decay into four leptons. 
Measurements of radiative $Z$ decay~\cite{Aad:2014wra} probe this final state,
with the best acceptance attained in the $4\mu$ channel.
The resulting limit is unrelated to the coupling of the $Z'$ to the electrons. It is not surprising that the bounds
one gets from this process are generally strong, 
constraining values of $\gZp \lesssim 10^{-2}$,
and for high $Z'$ masses this search is superior to the search that we propose. However, 
as one lowers the mass of the $Z'$, the search for the muons starts suffering from poor 
acceptances, as Ref.~\cite{Aad:2014wra} vetoes events with dileptons below 5 GeV to
suppress charmonium background.
Simultaneously, the BR of the exotic decay mode $Z\to Z' \gamma$ grows, and as we will see 
in Sec.~\ref{sec:exoticdecays}, becomes more sensitive than the four muon search. 

One can also consider the LHCb light $Z'$ search~\cite{Aaij:2017rft},  
which focuses on a light $Z'$ that is directly produced in the LHC $pp$
collisions and further decays into $\mu+ \mu^-$. No associated production is considered and 
no other decay channels except the dimuon decay modes. 
We also rescale the bound from this search 
and present it together with our projected reach on Fig.~\ref{fig:searchreach}. 
Not surprisingly the bound on lepton-specific models 
 is again relatively weak and comparable to the one that
is expected from the search of~\cite{Hoenig:2014dsa}, namely of order $\gZp \lesssim 10^{-1}$.   

Finally we comment on the flavor bounds. As we have noticed in 
Sec.~\ref{sec:models}, none of our $Z'$s is expected to \emph{mediate} FCNCs. 
However, the \emph{emission} of a light $Z'$ can give 
an important contributions to FCNC processes that are mediated by 
$W$ bosons. In this sense all of the processes that we are analyzing here, are by 
definition minimally flavor violating in the quark sector, and nonetheless the effect can be appreciable. Most of the relevant  processes
have been analyzed already in  Ref.~\cite{Dror:2017nsg} and we will rely on the results 
of this paper. For a light $Z'$, the dominant FCNC processes correspond to effective two-loop diagrams, such that the $Z'$ is emitted from a virtual $W$; note that this is true 
even if the quarks have charge under $U(1)'$, which would give FCNCs arising at one loop but
relatively suppressed by $m_{Z'} / m_t$.
The most stringent constraints on this scenario come from the $B \to K Z' $ 
and, to a lesser extent, $B \to \pi Z' $ decays. 
There are additional bounds from the $K \to \pi Z'$
decays are not relevant in our mass range of interest $m_{Z'} > 1$~GeV.  

Ref.~\cite{Dror:2017nsg} puts a bound of order $\gZp \sim 10^{-2}$--$10^{-3}$ 
in the case where the $Z'$ decays invisibly in the relevant mass range. For our light $Z'$
models where the dominant decays are to leptons, we are instead interested in muonic
decays. In this channel, the comparison 
to data is more laborious (see~\cite{Aaij:2012vr} for an 
experimental reference) due to the absence of the theoretical prediction and the necessity 
to compare bin-by-bin our predictions to the LHCb results. We prefer not to perform 
a full recast of this search in this paper, though the bound should not 
be very different from the one reported on the invisible decays in~\cite{Dror:2017nsg}.
Parenthetically, we also note that these constraints can be further attenuated 
in right-handed lepton number models, because, as we have explained, 
the dominant contribution to the process comes from emission of a $Z'$ from a 
virtual $W$ through the anomaly. Evidently, if we gauge the RH lepton number, the corresponding one-loop 
coupling is absent.

\section{Exotic Z Decays at the LHC}
\label{sec:exoticdecays}

In this section we describe the LHC analysis that we propose and roughly estimate
its expected sensitivity to a light $Z'$ in exotic $Z$ decays. In order to avoid 
large QCD backgrounds, we will further concentrate on the leptonic decays 
of the $Z'$, hence the choice of our benchmark models in Sec.~\ref{sec:models}.

\begin{figure}
	\centering
	\includegraphics[width=.49\textwidth]{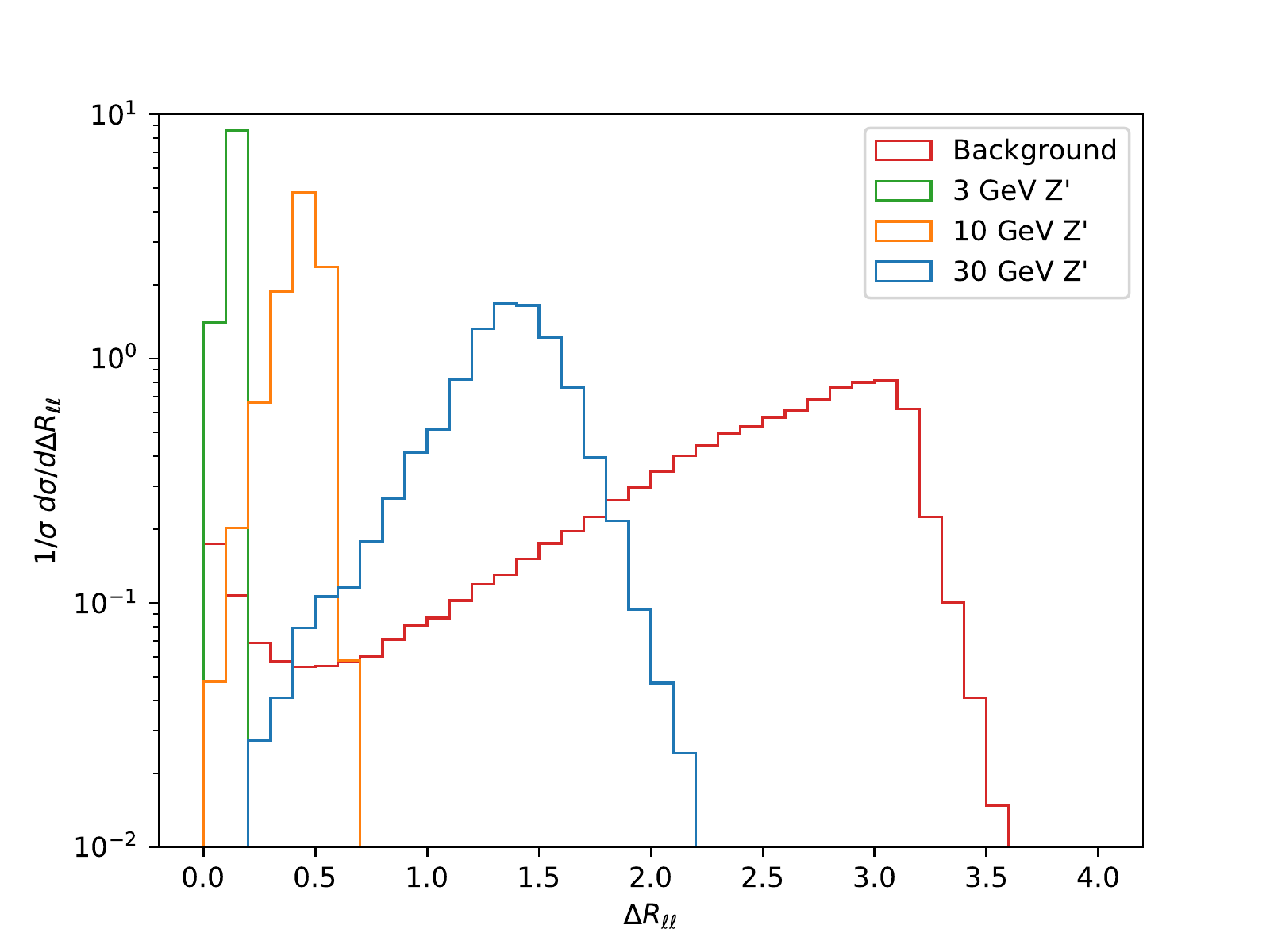}
	\includegraphics[width=.49\textwidth]{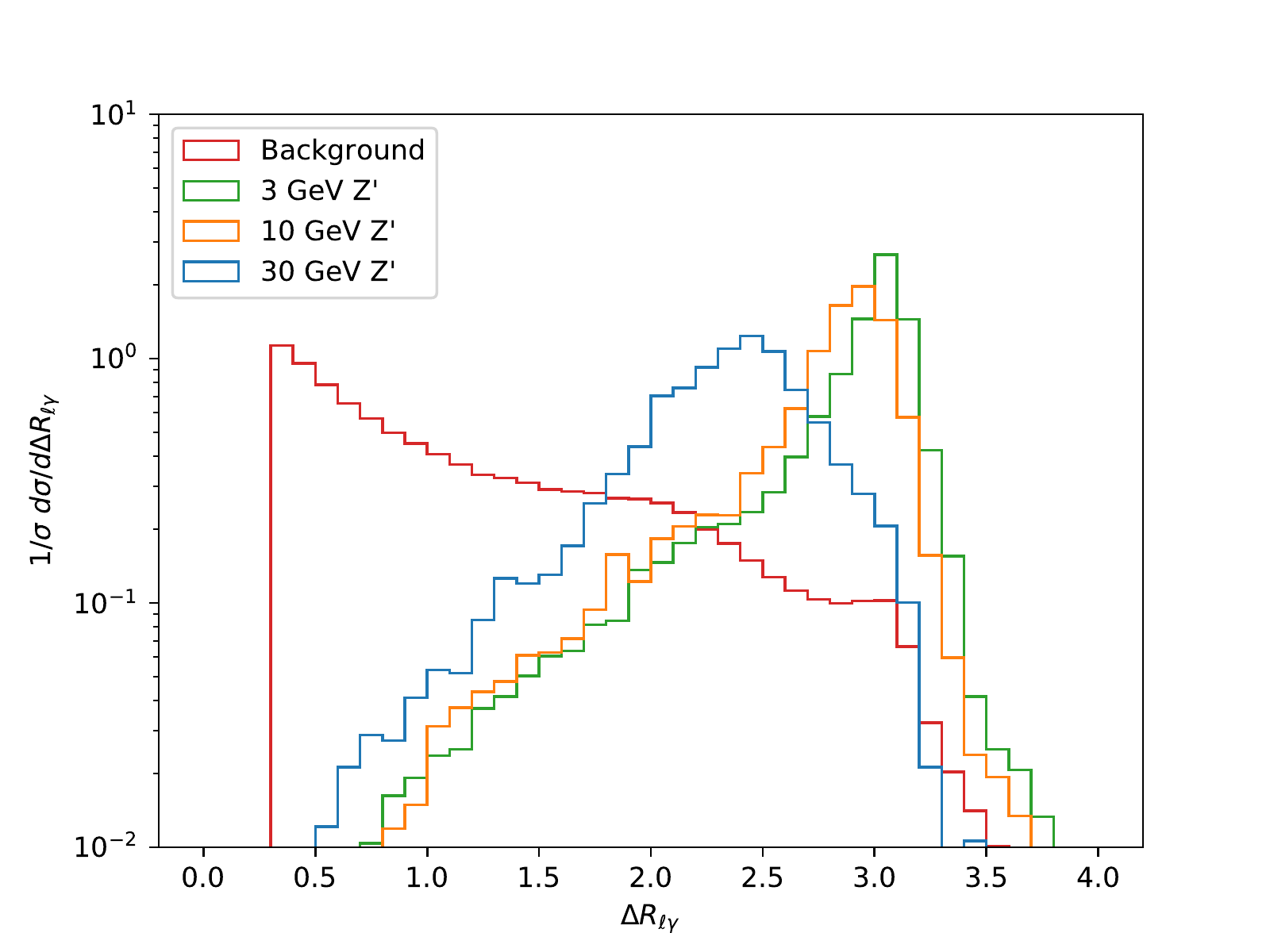}
	\caption{The separation $\Delta R$ between the leptons (left) and between 
		the photon and the nearest lepton 
		(right) as a function of the $Z'$ mass in the signal events, 
		compared to the background. All the distributions are normalized. 
}
	\label{fig:kinematics}
\end{figure}

The signature that we focus on in this study is $Z \to Z' \gamma,\ Z' \to \ell^+ \ell^-$ such that 
the invariant mass of 
the three final state objects reconstructs the $Z$ mass. The dominant background is 
leptonic $Z$ decay with a photon radiated off of a final state lepton. 
In our simulations we also generate the $\gamma^*/Z^* +  \gamma$ background where the real photon comes 
from initial state radiation, but this process is subdominant to the three-body $Z$ decays. 

While the background is by no means small, it has a very different geometry from the signal 
events. In the signal events the $Z'$ and the photon are back-to-back in the $Z$ rest frame, and 
because the boost of the $Z$ in the vast majority of the events is moderate, the picture in the lab frame 
is not vastly different. Moreover, because a light $Z'$ tends to be somewhat boosted, the leptons from its decay
will mostly follow the direction of the $Z'$. This behavior becomes more prominent for increasingly light $Z'$s. 
Therefore, in the bulk of the signal events the photon and leptons will be in different 
detector hemispheres. In the background events, however, the photons typically 
have small $p_T$ and are collinear with one of the leptons. We illustrate these properties 
of the signal and background,
showing the distribution of the angular separation between the photon and the 
closest lepton $\Delta R_{\ell\gamma}$ in Fig.~\ref{fig:kinematics}.\footnote{Both the signal and 
background events are simulated at leading 
order with \texttt{MadGraph 5}~\cite{Alwall:2014hca} and further showered and 
hadronized with \texttt{Pythia 6}~\cite{Sjostrand:2006za}, with matching up to one additional jet. No K-factors have been applied. 
The detector simulation has been performed with \texttt{Delphes 3}~\cite{deFavereau:2013fsa}.} The shoulder in the background $\Delta R_{\ell\gamma}$ distribution is due to $Z\gamma$ production, whose kinematics are more similar to our signal process than radiative $Z$ decay. Note, however, that the invariant mass 
of the $\ell^+\ell^-\gamma$ system usually 
significantly exceeds the $Z$ mass when the $Z$ recoils
against a hard photon.

Another kinematic variable that is relevant for the search is the angular distance between the two 
leptons $\Delta R_{\ell\ell}$. This variable is strongly correlated with 
$\Delta R_{\ell\gamma}$. 
In the background events the photon is usually soft and 
therefore the leptons are almost back-to-back and are very well separated 
from one another. In the signal events the leptons are coming from the decays 
of the $Z'$, which is again boosted if $\mZp \ll m_Z$. Therefore the signal leptons 
will often be spatially close to one another and even collimated in the extreme case of very 
light $Z'$, leading to ``lepton jets''. On general grounds we expect
\beq 
\Delta R_{\ell \ell} \sim \frac{4 m_Z m_{Z'}}{(m_Z^2 - m_{Z'}^2)}~. 
\eeq 
For example, for $\mZp = 15$~GeV,
$\Delta R_{\ell \ell} \sim 0.3$. This point is also illustrated 
in Fig.~\ref{fig:kinematics}. Again, $Z \gamma$ production leads to a small peak in the background at low $\Delta R_{\ell \ell}$. For any analysis requiring isolated leptons, however, this feature is irrelevant.

Finally we notice one more important feature of the background that will further 
largely determine our search strategy. After we impose an isolation cut on the 
photon (here we use the standard criterion of
$\Delta R > 0.3$ between 
the photon and any other reconstructed object in the event), the background is 
\emph{not} a smoothly falling function of the dilepton mass $m_{\ell\ell}$. In fact the cut on the 
spatial distance between the photon and the lepton introduces a hidden scale 
into the problem, because the the probability to emit an extra photon is 
proportional to the Sudakov double logarithm $\log^2 \delta p^2$, where $\delta p$ is the
change in the momentum of the emitting particle due to the bremsstrahlung.
Therefore the angular cut necessarily
translates into the minimal energy fraction that the emitted photon can carry, 
and the corresponding maximal dilepton invariant mass. 
For a $\Delta R > 0.3$ cut one finds a broad 
hump in the background $dN/dm_{\ell\ell}$ distribution, around 50~GeV. We emphasize 
again, that this is an expected feature of the background, carved by our cuts.  This
was also discussed in detail in
a recent experimental analysis~\cite{Khachatryan:2015rja}, although their feature was much 
 closer to the $Z$ mass due to much looser angular cuts than we propose. Clearly, as 
 we tighten the cut on the $\Delta R_{\ell \gamma}$ (or, alternatively, $\Delta R_{\ell \ell}$)
 the hump moves further to lower invariant dilepton masses, but it is always present. 
 
 \begin{figure}
 	\centering
 	\includegraphics[width=0.49\textwidth]{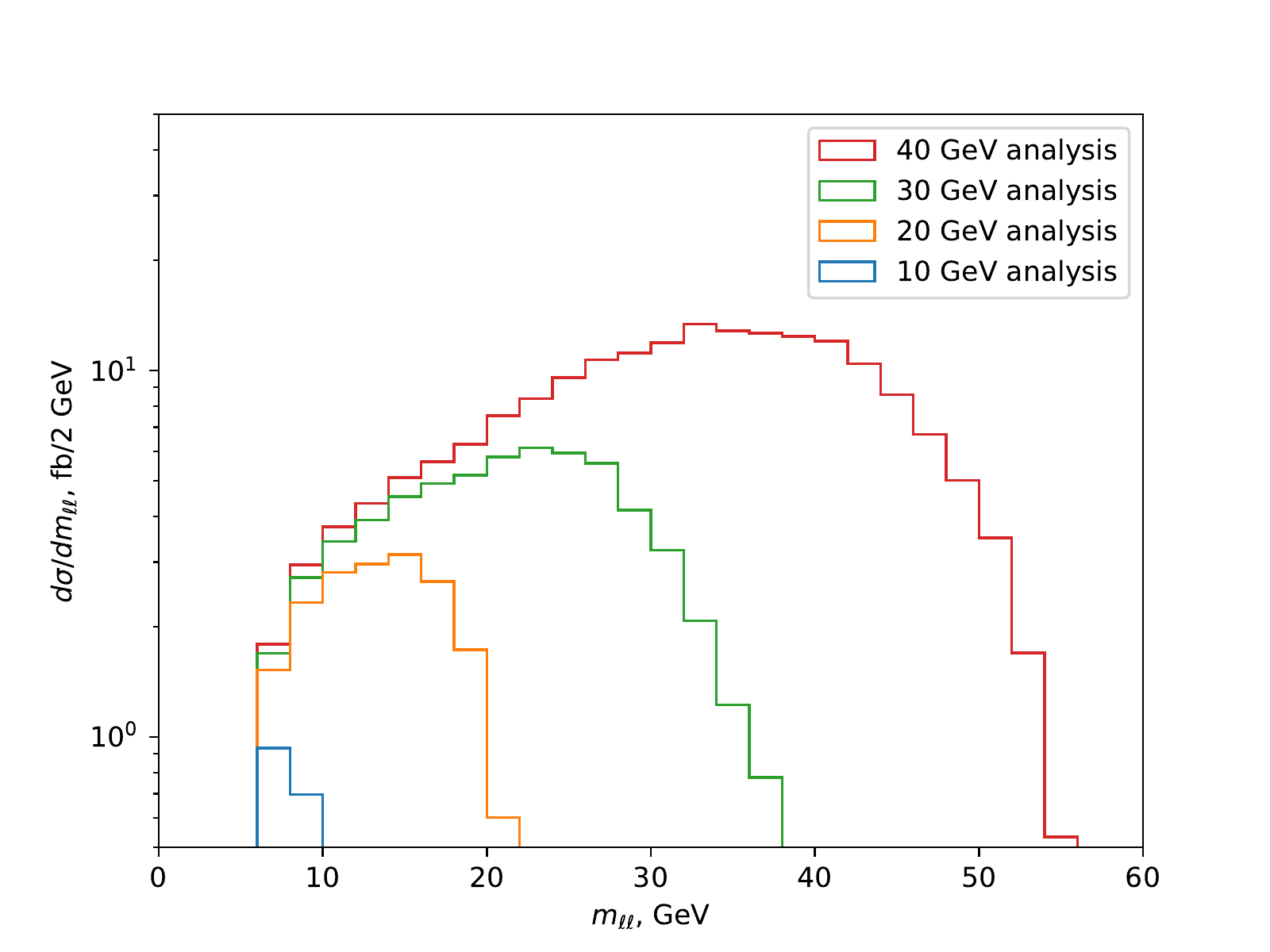}
 	\caption{Dilepton mass distribution for the $\ell\ell\gamma$ background. All cuts in the
 		text, except for the $m_{\ell\ell}$ resonance requirement,
 		 have been imposed, for various
 		target $Z'$ masses.}
 	\label{fig:mll}
 \end{figure}
 
 Because we suggest a search for a low mass resonance, bump hunting on top of the low-mass
 $m_{\ell\ell}$ background is a logical strategy.
 That said, the feature that we have just described will somewhat hamper these attempts. 
 Not only does the background have a non-trivial feature, resulting from our own cuts, but
 the signal can be close to the background feature itself, depending
 on the $Z'$ mass (although, for most of the target masses this is clearly not the case, as
 one can see from Fig.~\ref{fig:mll}, and the situation improves as we are going to lower 
 masses.) 
 Therefore, rather than fitting the overall dilepton mass distribution,
 we choose to impose angular cuts for a given $Z'$ mass and then look for a dilepton resonance.
 For the background, we expect that we can largely rely on the theoretical prediction, given that 
 the differential $Z$ production is theoretically known up to NNLO order, both 
 singly-produced~\cite{Melnikov:2006kv,Catani:2009sm} and in conjunction with 
 a photon~\cite{Grazzini:2015nwa}.   
 
 We proceed to perform a simple analysis for a given $Z'$ mass with the following cuts:
 \begin{itemize}
 \item Exactly two leptons with $p_T > 25, 10$~GeV and $|\eta| < 2.5$
 \item Exactly one photon with $p_T > 25$~GeV, $|\eta| < 2.5$
 \item $m_{\ell\ell\gamma} = m_Z \pm 5$~GeV
 \item $m_{\ell\ell} = m_{Z'} \pm 1$~GeV
 \item $\Delta R_{\ell\ell} < 4 m_Z m_{Z'} / (m_Z^2 - m_{Z'}^2)$
 \item $\Delta R_{\ell \gamma} > \pi - 4 m_Z m_{Z'} / (m_Z^2 - m_{Z'}^2)$
 \end{itemize}
 Throughout, we impose a minimum separation between physics objects $\Delta R_{\ell\ell}, 
 \Delta R_{\ell \gamma} > 0.3$. For $m_{Z'} \gtrsim 10$~GeV, the leptons are sufficiently well separated from each other that such isolation cuts do not significantly affect the acceptance.
 
 The impact of the angular cuts on the background dilepton mass distribution is shown
 explicitly  in 
 Fig.~\ref{fig:mll}. This plot shows the $m_{\ell\ell}$ distribution for the $\ell\ell\gamma$ 
 background after all of the cuts above except for the $m_{\ell\ell}$ requirement have been 
 imposed, with the angular cuts chosen to target a selection of $Z'$ masses. Clearly the 
 background is shaped by the analysis cuts, and is neither a smoothly falling function of the
 dilepton mass nor peaked at the $Z$ mass. The latter behavior  would be 
 approached if very soft photons aligned with the outgoing leptons were allowed by the cuts.
 
 For resonances below approximately 10 GeV, the decay products of the $Z'$ are collimated and a lepton-jet 
 search becomes appropriate. To avoid photon fakes, we do not consider 
 the electrons in this mass range and search
 for two muons within $\Delta R_{\ell\ell} < 0.5$~\cite{Aad:2015sms}, as well as a well-separated 
 photon $\Delta R_{\ell \gamma} > \pi - 0.5$. At such low masses, meson resonances become 
 important background sources, and so we take a tighter dilepton mass window $m_{\ell\ell} = m_{Z'} \pm 20$~MeV~\cite{Aad:2014rra}.
 
\begin{figure}
	\centering
	\includegraphics[width=.49\textwidth]{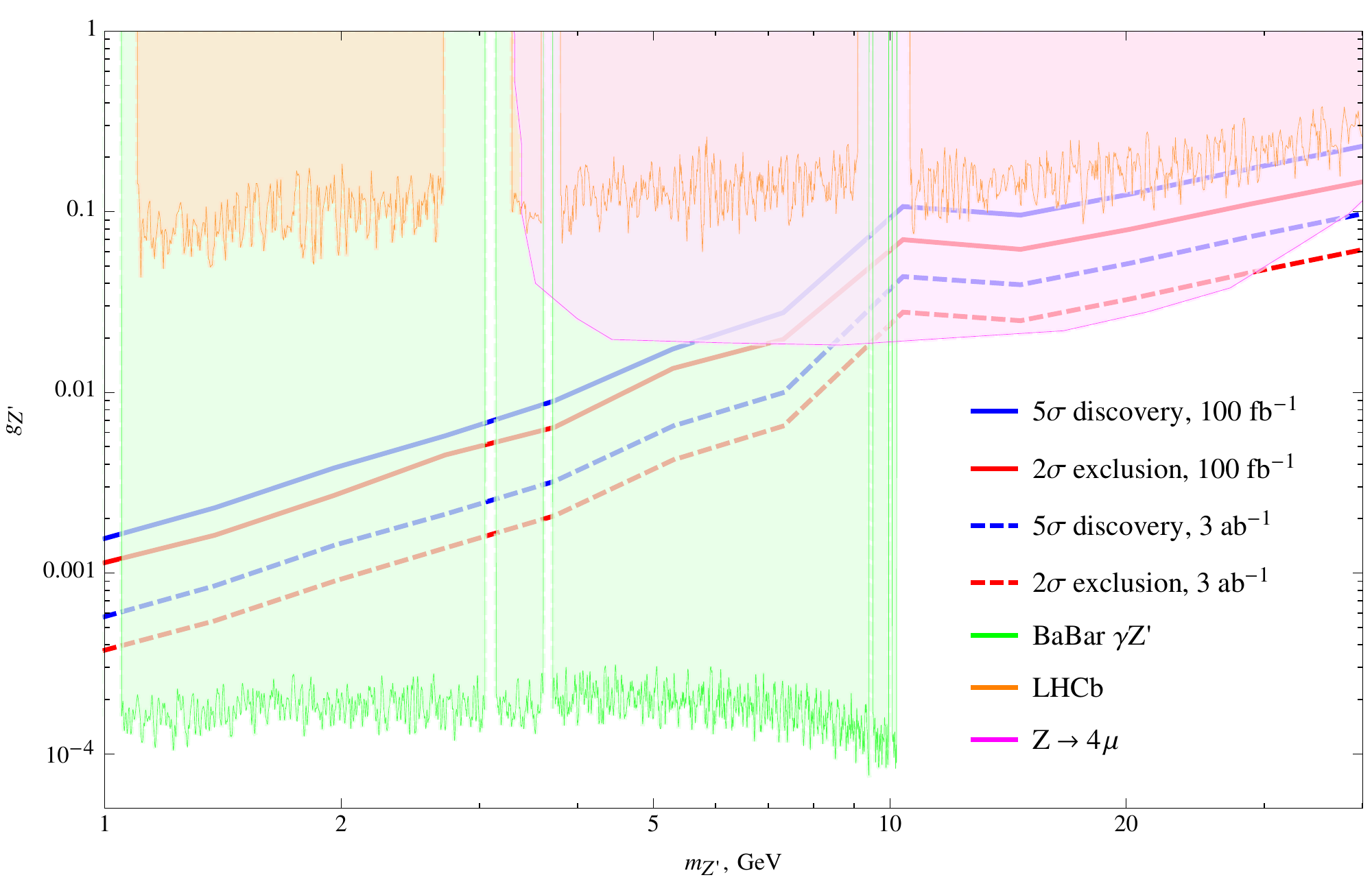}
	\includegraphics[width=.49\textwidth]{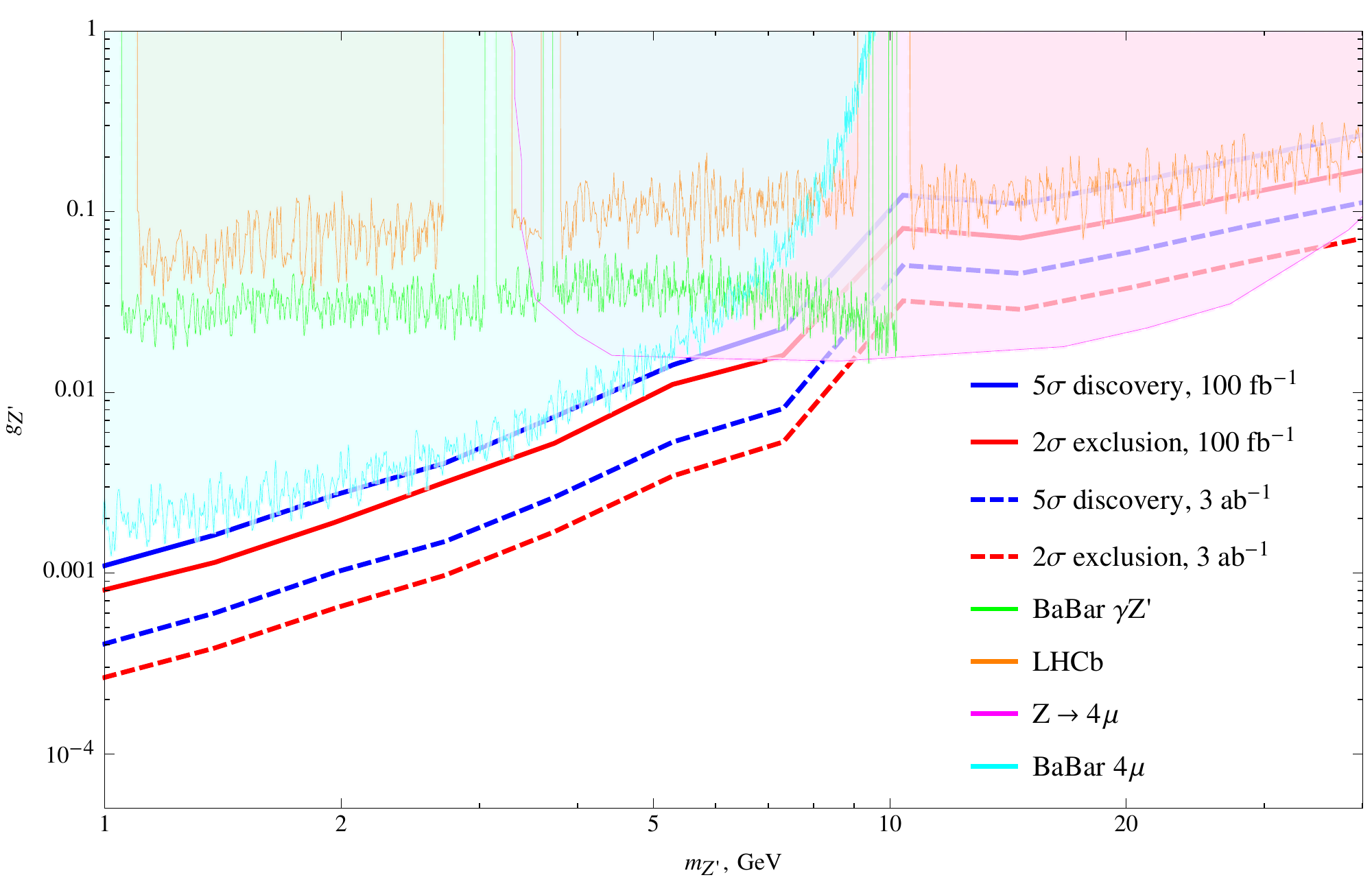}
	\caption{Exclusion ($2\sigma$) and discovery ($5\sigma)$ contours for the $\ell\ell\gamma$
	searches described in the text. The lepton number model of Sec.~\ref{sec:models} is
	assumed, with $\cA_{Z'BB} = - \cA_{Z'WW} = - N_g/2$. The shaded regions indicate the (suggested)
bounds from other experiments. The picture on the left hand side concerns with the lepton number 
theories, that couple indiscriminatingly  to all the lepton generations. The picture on the RH side 
shows the situation, assuming that the $Z'$ couples only to the second and the third generations 
of the leptons ($\mu + \tau$). }
\label{fig:searchreach}
\end{figure}

 We present the expected reach of the searches 
 that we propose in Fig.~\ref{fig:searchreach}, for both a leptophilic
 $Z'$ as well as a $L_\mu + L_\tau$ gauge boson. For comparison, the dominant constraints from
 Sec.~\ref{sec:constraints} are also shown. We see that for direct couplings of the electron 
 to the $Z'$, the BaBar limit from $e^+ e^- \to Z' \gamma$ is quite stringent for $Z'$ masses 
 that are kinematically accessible, and at higher masses the existing LHC $Z \to 4 \mu$ search 
 already excludes the parameter space that would be probed by our analysis. Nevertheless, we
 emphasize that rare $Z$ decays constitute an independent check of the mixed 
 anomaly coefficients once a $Z'$ is discovered, without requiring the observation of all of
 the couplings in the low-energy anomalous EFT. The lack of such decays could be construed as
 evidence for an anomaly-free gauge group with the 
 SM matter field content, such as $B - L$, sequential hypercharge or $L_\mu - L_\tau$.
 
 When $L_\mu + L_\tau$ is gauged, our proposed lepton jet search is already competitive with 
 only 100~fb$^{-1}$ of LHC data, which should be achievable next year. 
 Belle II~\cite{Aushev:2010bq}, with a 
 large increase in integrated luminosity over BaBar, 
 would be able to improve the BaBar $e^+ e^- \to 4 \mu$ bound in the 
 light $Z'$ region. For intermediate $Z'$ masses in the 5--10 GeV range, several existing 
 searches show roughly similar sensitivity, including both BaBar searches for hidden gauge
 bosons and LHC rare $Z$ decays. With more data, the lepton jet analysis proposed in this 
 work should be able to outperform existing analyses, though a full study with systematics
 would be worthwhile. At higher $Z'$ masses, the $Z \to 4 \mu$ search is expected to remain
 dominant, although, as we have pointed out, the searches for the exotic $Z$ decays
 can carry valuable information.

\section{Conclusions}
\label{sec:conc}

Because generic $U(1)'$ theories contain mixed anomalies with the SM at collider energy scales, most gauged Abelian extensions of the SM include Chern-Simons 
terms coupling the new gauge boson to a pair of SM bosons. 
Knowing only the charges of the low energy fermionic content of the theory under a new $U(1)'$ allows one to determine these Chern-Simons terms unambiguously 
up to fermion mass effects, 
assuming the absence of new sources of electroweak symmetry breaking.

In this study we have examined the exotic decay mode $Z \to Z' \gamma$ induced 
by such anomalous terms. Because the associated operators are simply higher dimensional operators of an EFT, their influence rises with $m_Z / m_{Z'}$. In particular, the BR for the exotic decay can be detectable at the LHC for a sufficiently light $Z'$. We have demonstrated a simple analysis that could be used to search for exotic $Z$ decays, including using lepton jets in the region where the decay products of the $Z'$ are collimated. The resulting limits can be competitive with existing searches, owing to the large $Z$ production cross section at the LHC.

While our toy search relied on the existence of a leptonic $Z'$ coupling, $Z'$ production in $Z$ decays does not depend on any particular interaction other than the anomalous Chern-Simons terms. It is thus conceivable that a low mass dijet resonance search could also be performed in events with photons. Because the dijet resonance would be boosted, it would be crucial to manage the QCD background well and have a good understanding of substructure variables.


No matter how a new gauge boson is discovered, the searches for rare decays which we have outlined can be instrumental in revealing the underlying structure of new physics. In a fashion completely orthogonal to traditional dijet or dilepton resonance analyses, gauge boson resonance searches offer insight into the anomaly coefficients of a gauge theory at one or more energy scales. Such information constitutes an independent probe of the physics associated with a new gauge symmetry that could lie just beyond the SM.

\acknowledgments

We are grateful to Bertrand Echenard, Tobias Golling, and Eder Izaguirre for useful discussions, to Maxim Pospelov for helpful discussions and comments on the manuscript, and to the anonymous referee for pointing out an error in the first version of this work. 
The work of AI is supported in part by the U.S. Department of Energy under grant No.~DE-SC0015634, 
and in part by PITT PACC.

\bibliographystyle{JHEP}
\bibliography{dmanomaly}

\end{document}